\documentclass[9pt,conference]{IEEEtran}

\newcommand{\papername}{\textit{TinyCall} }

\usepackage{graphicx}
\usepackage{subcaption}
\usepackage{booktabs}
\usepackage{balance}
\usepackage{amsmath}
\usepackage{adjustbox}
\usepackage{array}
\usepackage{float}
\usepackage{makecell}
\usepackage{pifont}
\usepackage{multirow}
\usepackage[dvipsnames]{xcolor}
\usepackage{microtype}
\usepackage{enumitem}
\usepackage{tikz}
\usepackage{xspace}
\usepackage{dcase2026}

\graphicspath{{figures/}{images/}{../images/}}
\setlist{nosep,leftmargin=*}

\newcommand{\aritrik}[1]{\textcolor{black}{#1}}

\newcommand{\squishlist}{
  \begin{list}{$\bullet$}{
    \setlength{\leftmargin}{0.15in}
  }
}

\newcommand{\squishend}{
  \end{list}
}

\title{Narrowband Voice Communication Using Streaming Neural Compression}

\name{Dahong Luo, Anannya Trehan, Aritrik Ghosh, Nirupam Roy\vspace{-0.15in}}
\address{University of Maryland, College Park}

\begin{document}
\maketitle

\begin{abstract}
Low-bitrate speech communication on resource-constrained edge devices
remains challenging due to stringent computational, memory, and bandwidth
constraints. We present \papername, a lightweight neural audio codec
designed for real-time speech communication on low-power platforms such
as the ESP32 microcontroller and Raspberry Pi. The proposed system
targets emergency communication and other bandwidth-limited scenarios
while preserving speech intelligibility, speaker identity, and vocal
expressiveness.
To enable efficient deployment, we propose a minimal neural audio codec
architecture together with a framework for converting a causally trained
codec into a truly streamable codec through pseudo-lookahead decoding and
decoder-input caching. We further replace conventional residual vector
quantization (RVQ) with Residual Finite Scalar Quantization (RFSQ) to
reduce inference complexity on edge processors and employ a progressive
three-stage training strategy for stable optimization under latent
quantization. An MFCC-based perceptual loss encourages preservation of
speaker characteristics, including harmonic structure and vocal timbre.
Experimental results demonstrate real-time operation on a Raspberry
Pi~3 while achieving intelligible speech reconstruction at bitrates as
low as 2.3~kbps. The proposed approach demonstrates that practical
neural speech communication is feasible on highly resource-constrained
edge devices.
\end{abstract}

\begin{IEEEkeywords}
Neural speech coding, Streaming audio processing, Audio codecs, Speech reconstruction.
\end{IEEEkeywords}

\section{Introduction}
\vspace{-0.05in}
For decades, voice communication has been synonymous with phones. We envision a different future: voice connectivity becoming a fundamental capability of embedded computing itself. Just as wireless sensing, localization, and internet connectivity have migrated into tiny low-power devices, voice communication could become an always-available primitive integrated into keychains, wearables, safety tags, asset trackers, and other embedded systems. Unlike text messages, alerts, or predefined status notifications, speech enables fast and natural interaction while conveying emotion, urgency, and speaker identity—qualities that make human communication personal, recognizable, and effective. Such ubiquitous voice connectivity could enable applications ranging from child and elderly safety to remote-worker communication and interactive emergency beacons, while providing a natural communication interface for future IoT and human-centered computing systems. Realizing this vision, however, requires voice communication to operate within resource budgets that are orders of magnitude smaller than those assumed by traditional systems.

To retain the simplicity and longevity of embedded endpoints, such devices must run on low-cost, low-power wireless MCUs that provide only modest resources, typically 1--1.5 MB of non-volatile storage and 128--512 KB of RAM (e.g., Nordic nRF54L15, ST STM32WBA52, Silicon Labs EFR32BG24, and Espressif ESP32-S3)~\cite{nrf54l15,stm32wba52,efr32bg24,esp32s3}. Communication links are similarly constrained: low-power wide-area networks trade throughput for range and energy efficiency~\cite{raza2017lpwan}, with LoRaWAN supporting roughly 0.3--50 kbps, NB-IoT providing 10--100 kbps depending on coverage conditions, and Sigfox-class ultra-narrowband systems being too slow for live audio altogether~\cite{lorawanSpec,nbiotRates,sigfoxRate}. At the same time, a practical voice communication system must preserve more than lexical intelligibility. Speech conveys speaker identity and prosodic cues associated with emotion and urgency~\cite{helander2007prosody,scherer2004prosody}, enabling listeners to recognize who is speaking and infer their affective state. Fitting intelligible, expressive, and recognizable speech within these stringent compute, memory, and bandwidth budgets remains a significant challenge.

\aritrik{Existing speech codecs meet these requirements only in part. Opus and Codec2 are strong baselines for interactive and bandwidth-constrained
voice~\cite{opus,codec2}, but Opus degrades sharply below roughly
6 kbps, and Codec2, while reaching 0.7-3.2~kbps, sacrifices
naturalness and speaker identity. Neural codecs such as SoundStream,
EnCodec, AudioDec, and DAC deliver high quality at low
bitrates~\cite{soundstream,encodec,audiodec,dac}, but with millions of
parameters and residual vector quantization (RVQ) search, their
footprints target server or smartphone-class hardware. This gap
motivates our design target: a streaming speech codec operating sub 2.5 kbps, with fewer than 1 million parameters, fixed-size
streaming state, and real-time execution on an edge-class processor.}

\aritrik{We present \papername, a lightweight streaming neural codec for this setting. To keep inference practical on modest hardware, \papername
replaces RVQ---whose repeated nearest-neighbor searches across
multiple codebooks dominate quantization cost---with Residual Finite
Scalar Quantization (RFSQ), which rounds each latent dimension to
fixed scalar levels and eliminates the search entirely~\cite{fsq}.
Because RFSQ stores no learned codebooks, it also contributes nothing
to model size or decoder state, a direct saving under the memory
budgets above. A progressive training strategy further stabilizes
reconstruction quality at the low end of the bitrate range.}

\aritrik{\papername reconstructs speech by predicting a time-frequency
representation and synthesizing the waveform via inverse short-time
Fourier transform (iSTFT). This spectral formulation lightens the
neural decoder but complicates streaming, since with frame overlap, reconstructing frame
$t$ requires information from frame $t+1$ that has not yet arrived.
% We resolve this with pseudo-lookahead decoding and a small
% latent-state cache: when the latent representation for frame $t+1$
% arrives, the decoder reconstructs and releases frame $t$, whose
% neighboring context is now complete. This preserves streaming
% operation at the cost of a one-frame algorithmic delay. Because the
% cache holds only a fixed frame of latent state, decoder memory
% remains fixed regardless of utterance length.
We resolve this with a pseudo-lookahead decoding strategy and a small latent-state cache, which preserves streaming operation at the cost of a one-frame algorithmic delay.}
\aritrik{Our contributions in this paper can be summarized as follows.}
\begin{itemize}
    \item \aritrik{A compact causal neural codec sub 3-kbps voice communication on resource-constrained edge processors.}
    \vspace{-0.12in}
    
    \aritrik{\item An RFSQ bottleneck and progressive training strategy for
    efficient low-bitrate speech reconstruction.}
    \vspace{-0.12in}
    
    \aritrik{\item Pseudo-lookahead decoding and latent caching that make
    iSTFT-based reconstruction streamable.}
    \vspace{-0.12in}

    \aritrik{\item Implementation on embedded device for real-time intelligible speech encoding and reconstruction.}
    %at source bitrates as low as 2.3 kbps.}
\end{itemize}

\section{Related Work}
%\vspace{-0.05in}

\subsection{Neural Audio Codecs}
\vspace{-0.05in}
%\noindent {\bf Neural Audio Codecs:} Early neural audio codecs, including SoundStream~\cite{soundstream}
and EnCodec~\cite{encodec}, established end-to-end compression
pipelines that reconstruct raw time-domain waveforms from discrete
latent tokens. While effective, time-domain decoding often requires
adversarial training and deep upsampling stacks.
Recent codecs reduce this cost by moving reconstruction to the
frequency domain. SpectroStream~\cite{li2025spectrostreamversatileneuralcodec},
STFTCodec~\cite{Feng2025STFTCodecHA}, and APCodec~\cite{APcodec}
predict spectral features instead of raw samples, reducing decoder
complexity for high-sample-rate audio. Vocos~\cite{Vocos} follows this
direction by predicting complex spectrograms and using deterministic
iSTFT synthesis for waveform reconstruction. We adopt this
spectral-decoding approach for its computational efficiency and audio
fidelity.
%\vspace{0.03in}

\subsection{Quantization Techniques}
\vspace{-0.05in}
%\noindent {\bf Quantization Techniques: }
Most modern neural codecs use Residual Vector Quantization
(RVQ)~\cite{soundstream} to map continuous latents into discrete
bitstreams. Although effective, RVQ requires sequential codebook
searches, can suffer from codebook collapse, and often needs auxiliary
losses to stabilize codebook usage. Finite Scalar Quantization
(FSQ)~\cite{fsq} simplifies this process by projecting each latent
dimension onto a fixed bounded scalar grid, eliminating learned
codebooks and improving code utilization.
However, directly stacking FSQ stages can lead to residual magnitude
decay: later stages receive progressively weaker residuals and may
underuse their available codes. Robust Residual FSQ
(RFSQ)~\cite{zhu2026robustresidualfinitescalar} addresses this with
learnable per-stage scaling and invertible normalization, preserving
residual capacity across stages without increasing the compressed
bitrate. We use RFSQ to obtain a stable, expressive bottleneck with
lower inference-time quantization overhead than RVQ.
%\vspace{0.03in}

\subsection{Streamable Audio Codecs}
\vspace{-0.05in}
%\noindent {\bf Streamable Audio Codecs: }
\aritrik{Low-latency neural codecs must support frame-by-frame execution with
bounded buffering, state, and synthesis delay. AudioDec~\cite{audiodec}
is an open-source streaming codec for real-time speech reconstruction,
while HILCodec~\cite{hilcodec} targets lightweight real-time audio
coding. Recent systems push this direction further: DualStream studies
ultra-low-delay neural audio coding with reduced computation
~\cite{dualstream}, and StreamCodec designs a fully causal MDCT-domain
codec for real-time communication~\cite{streamcodec}. These works show
that neural codecs can be made streamable, but practical deployment
still depends on the decoder schedule and memory footprint. This is
especially important for iSTFT-based decoders, where overlap-add
reconstruction depends on neighboring spectral frames. We address this
with pseudo-lookahead decoding and latent caching, enabling bounded
one-frame-delay streaming.}
%\vspace{0.03in}

%\subsection{Streamable Audio Codecs} Low-latency neural codecs such as AudioDec~\cite{audiodec} and HILCodec~\cite{hilcodec} are designed for real-time or streamable audio coding. These systems typically enforce architectural causality so that the model does not use future input during training or inference. However, causality alone does not guarantee efficient deployment on resource-constrained edge devices. A practical streaming codec must also manage historical context, per-frame buffering, memory movement, and decoder-side synthesis delay. Large receptive fields can still require substantial cached context and computation at every frame, increasing the real-time factor on low-power hardware. In our case, iSTFT-based decoding introduces an additional challenge because overlap-add reconstruction depends on neighboring spectral frames. We address this with pseudo-lookahead decoding and latent caching, enabling bounded one-frame-delay streaming.

 \subsection{Audio Communication on Edge Devices}
 \vspace{-0.05in}
%\noindent {\bf Audio Communication on Edge Devices: }
\aritrik{Prior work has explored audio communication on low-power and embedded devices. AudioCast shows that audio-broadcasting tags can
provide low-power connectivity for embedded systems~\cite{audiocast},
while backscatter-based systems demonstrate that acoustic
signals can be communicated with extremely lightweight back-scatter hardware
~\cite{fmbackscatter,pab}. More directly, ReSoNate studies audio
transmission over LPWANs using Codec-2 on embedded
platform, and VLoRA investigates real-time voice over a single LoRa
physical-layer channel for emergency communication~\cite{resonate,vlora}.
\papername is complementary it targets the source-coding layer needed to produce an intelligible and expressive live-voice stream small enough for constrained edge links while remaining practical for resource-limited processors.}
\section{Proposed Method}
%\vspace{-0.05in}

\subsection{\papername Codec Design}
\vspace{-0.05in}

The input waveform is first processed by an encoder based on the SEANet~\cite{SEANet} architecture, leveraging its residual convolutional blocks for highly efficient representation learning. To support real-time deployment, every convolutional layer in the encoder is modified to use causal one-sided padding. This strictly eliminates dependencies on future input samples, ensuring the output at each time step depends solely on current and previous inputs. With this configuration, the encoder requires 195k parameters.

The resulting sequence of latent vectors is then compressed using Residual Finite Scalar Quantization (RFSQ) before being transmitted. At the receiver, we replace the standard SEANet waveform decoder with a Vocos-style decoder (577k parameters), which incorporates the same causal padding constraints. This decoder predicts complex time-frequency spectrogram representations rather than raw audio, which are subsequently converted back into a time-domain waveform via a differentiable inverse short-time Fourier transform (iSTFT) head. This purely causal architecture serves as the foundation for the streamable decoding strategy detailed in Section \ref{streamble conversion}.

\begin{figure*}[t]
    \centering
    \includegraphics[width=\linewidth]{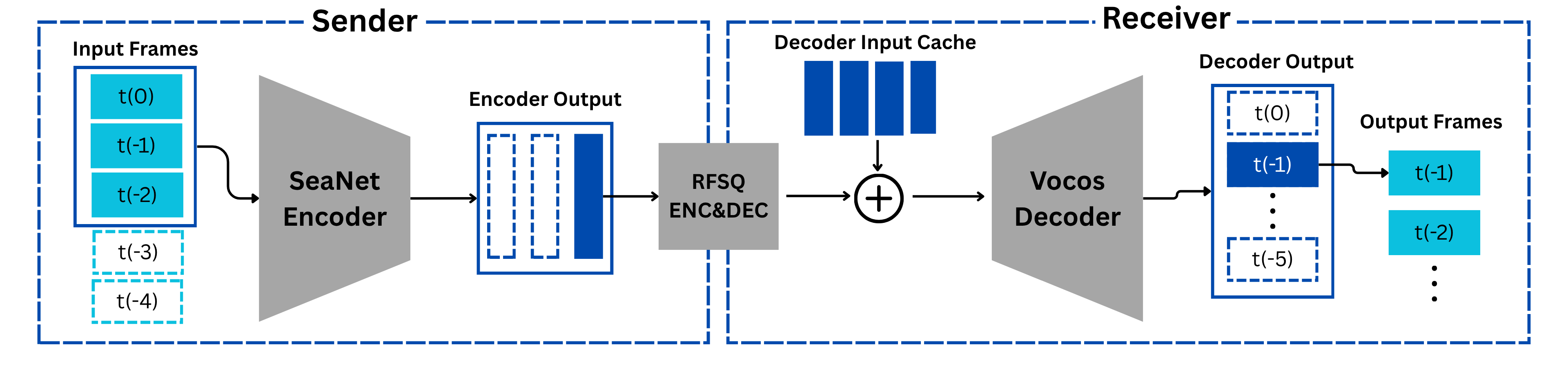}
    \vspace{-0.3in}
    \caption{Streaming inference pipeline of the proposed neural audio codec. Input audio is encoded into latent representations, compressed using RFSQ, and reconstructed through a decoder with latent input caching and pseudo-lookahead decoding to enable real-time streamable inference.}
    \vspace{-0.1in}
    \label{fig:overview}
\end{figure*}

Instead of the Residual Vector Quantization (RVQ) used by many existing neural audio codecs, the proposed model employs Residual Finite Scalar Quantization (RFSQ). During deployment on embedded processors, we observed that the repeated nearest-neighbor codebook searches required by RVQ dominate the inference cost, with quantization and dequantization alone consuming approximately 250\% of the available computation budget for real-time processing. RFSQ eliminates these lookup operations by independently quantizing each latent dimension using fixed scalar quantization levels, significantly reducing computational complexity while maintaining intelligibility and speaker identity at 2.3 kbps. 

Specifically, our RFSQ implementation utilizes four residual stages with non-uniform scalar quantization levels: [8, 8, 8, 4, 4, 4] for the first two stages, and [4, 4, 2, 2, 2, 2] for the final two. This non-uniform codebook sizing is designed so that the earlier stages encapsulate more critical signal information than the subsequent residuals. Consequently, the bits required per code evaluate to $\log_{2}(8^3 \cdot 4^3) = 15$ bits for each of the first two stages, and $\log_{2}(4^2 \cdot 2^4) = 8$ bits for the last two, yielding a total of 46 bits per frame. Operating at a 16 kHz sampling rate with a frame size of 320 samples, the model processes 50 frames per second, resulting in a highly efficient overall bitrate of 2.3 kbps.

\subsection{Causal-to-Streamable Conversion}
\label{streamble conversion}
\vspace{-0.05in}

Although many neural audio codecs are described as causal, causal inference alone does not guarantee its deployability in a real-time streaming system. Especially when a spectrogram-based decoder is employed, additional dependencies are introduced during waveform reconstruction that prevent straightforward frame-by-frame decoding.

The proposed codec employs a Vocos-style decoder followed by a differentiable inverse short-time Fourier transform (iSTFT) head. During training, the decoder predicts spectrogram frames for the entire utterance, and the waveform is reconstructed through the overlap-add operation of iSTFT. Consequently, the reconstructed waveform corresponding to frame $t$ is formed by the aggregation of neighboring spectrogram frames, including $t-1$, $t$, and $t+1$. While future frame information is readily available during offline training, it is unavailable during real-time inference. Directly reconstructing frame $t$ without access to frame $t+1$ results in incomplete overlap-add synthesis, producing distortion near frame boundaries.

To bridge this gap, we introduce a pseudo-lookahead decoding strategy that converts a causally trained codec into a truly streamable system with only a one-frame algorithmic delay. Instead of reconstructing the current frame immediately, the decoder predicts the waveform corresponding to frame $t-1$ after receiving the latent representation for frame $t$. At this point, all neighboring spectrogram information required for overlap-add reconstruction of frame $t-1$ has become available, allowing the decoder to accurately reproduce the same reconstruction process used during offline training.

To enable this fully streamable configuration, the system relies on two critical hyperparameters that govern temporal buffering: the input frame size and the decoder cache size. On the encoding side, the system maintains a queue of raw audio frames whose length is defined by the input frame size. This sliding window captures the necessary temporal context for the encoder's convolutional neural network (CNN) layers. While the encoder processes this entire buffer and outputs a sequence of latent codes corresponding to the input frame size, the streaming pipeline extracts the final code representing the current time step $t$ to the quantization bottleneck. On the decoding side, a similar buffering mechanism supports the pseudo-lookahead strategy. The decoder maintains a latent buffer governed by the decoder cache size. For each incoming quantized latent vector, the oldest entry is discarded, the newest is appended, and the resulting sequence is provided as the decoder input. In our implementation, the decoder cache size is set to five frames (the current latent representation plus the four preceding it). As the cache holds only a fixed frame of latent state, decoder memory remains fixed regardless of utterance length. Figure~\ref{fig:overview} presents an overview of the \papername architecture.

This approach preserves the computational efficiency of causal training while eliminating the reconstruction artifacts that arise from naively applying iSTFT synthesis in a streaming setting. More importantly, it enables models trained under standard causal conditions to be deployed as streamable neural audio codecs without modifying the training procedure or introducing future dependencies during encoding.

\subsection{Training Strategy}
\label{training strategy}
Training a lightweight neural audio codec with adversarial supervision and latent quantization from the beginning was found to produce unstable optimization. To improve convergence, we adopt a progressive three-stage training strategy that gradually introduces additional training objectives.

In the first stage, the encoder and decoder are trained using only reconstruction losses, allowing the model to learn a stable continuous latent representation. After 70k training steps, adversarial training is enabled by introducing the HiFiGAN Period Discriminator~\cite{kong2020hifigangenerativeadversarialnetworks} to improve perceptual speech quality.

The third stage begins after 250k training steps with the introduction of Residual Finite Scalar Quantization (RFSQ). Rather than immediately replacing the continuous latent representation, the model gradually transitions to the quantized representation using
\begin{equation}
    z_{\mathrm{mixed}} = (1-\alpha)z + \alpha z_q,
\end{equation}

where $z$ and $z_q$ denote the continuous and quantized latent representations, respectively, and $\alpha \in [0,1]$ is the transition rate. The transition progresses smoothly from 0 to 1 over 100k training steps, allowing the encoder to adapt to quantization error before relying entirely on quantized latents. Training then continues using fully quantized representations.

% The reconstruction objective combines an L1 waveform loss, a multi-scale STFT loss, and an MFCC-based perceptual loss. The MFCC loss encourages the preservation of speaker-dependent characteristics such as harmonic structure and vocal timbre. As illustrated by the spectrogram comparisons in Figure \ref{fig:mfcc_stft}, models trained with the MFCC objective exhibit sharper harmonic contours and fewer spectral artifacts compared to those relying solely on standard STFT and L1 losses. Finally, adversarial supervision from the HiFiGAN Period Discriminator further refines the perceptual quality of the generated waveform.

\begin{figure}
    \centering
    \includegraphics[width=1\linewidth]{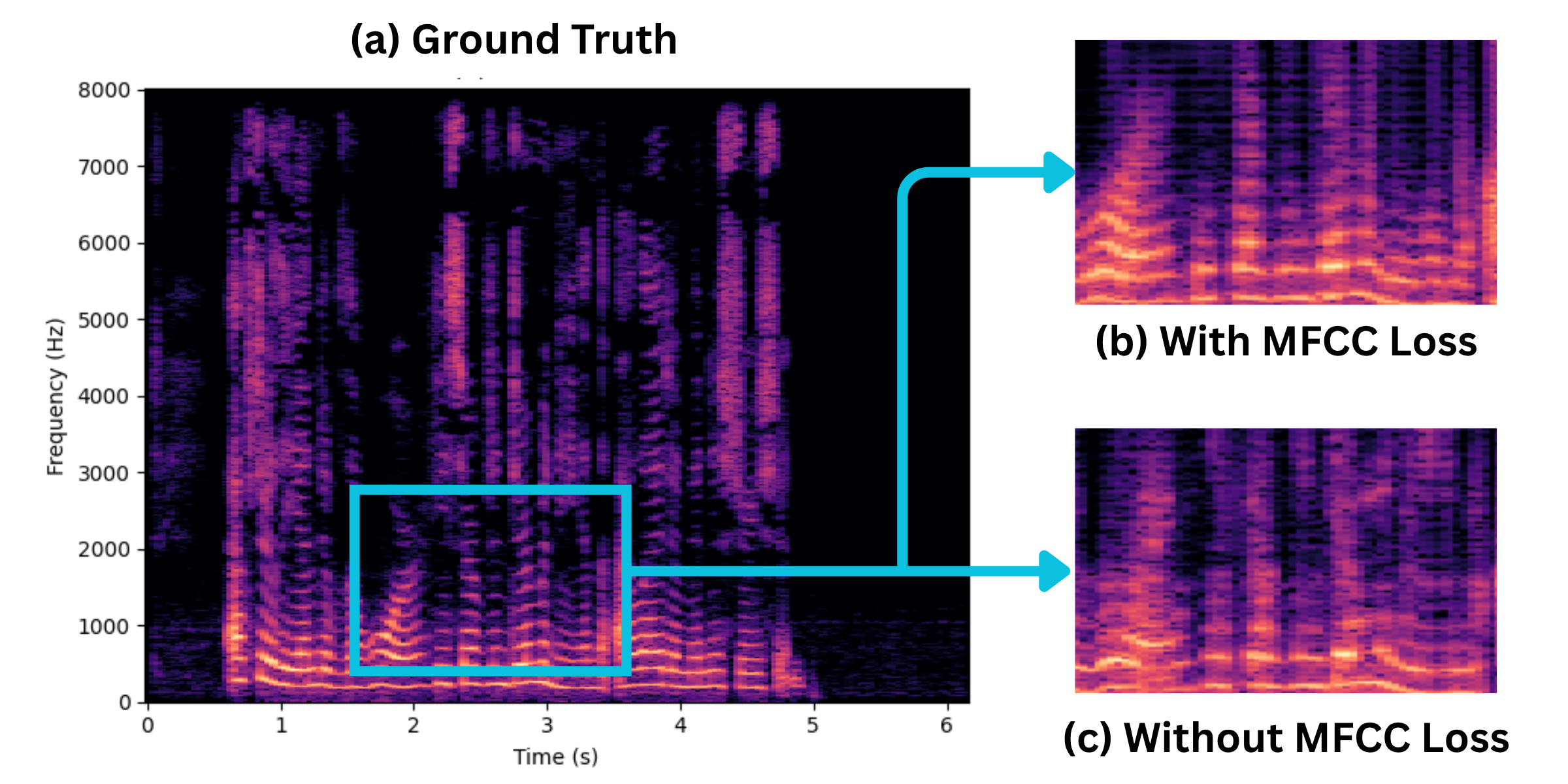}
    \vspace{-0.15in}
    \caption{Spectrogram comparison demonstrating the impact of MFCC-based perceptual loss. (a) Ground truth, alongside zoomed-in reconstructions from models trained (b) with and (c) without MFCC loss. The inclusion of the MFCC objective significantly improves harmonic preservation and reduces spectral artifacts in the highlighted region.}
    \label{fig:mfcc_stft}
    \vspace{-0.1in}
\end{figure}

The reconstruction objective combines an L1 waveform loss, a multi-scale STFT loss, and an MFCC-based perceptual loss:
\begin{equation}
    \mathcal{L}_{\text{MFCC}} = \left| \text{MFCC}(x) - \text{MFCC}(\hat{x}) \right|_1
\end{equation}
This formulation preserves speaker-dependent characteristics like harmonic structure and vocal timbre. As Figure \ref{fig:mfcc_stft} illustrates, models trained with $\mathcal{L}_{\text{MFCC}}$ exhibit sharper harmonic contours and fewer spectral artifacts than those relying solely on standard STFT and L1 losses. Finally, adversarial supervision from the HiFiGAN Period Discriminator further refines the generated waveform's perceptual quality.

\section{Evaluation}
%\vspace{-0.05in}

\subsection{Experimental Setup}
\vspace{-0.05in}

The proposed model is trained on the English subset of the Mozilla Common Voice corpus~\cite{CommonVoice}. Evaluation is conducted using a combination of the LibriSpeech~\cite{LibriSpeech} test set and EmoVoice-DB\cite{yang2025emovoicellmbasedemotionaltexttospeech}, allowing assessment on both neutral and emotionally expressive speech. LibriSpeech is used to evaluate general speech reconstruction performance, while EmoVoice-DB is included to assess the model's ability to preserve speaker characteristics and vocal expressiveness.

All models are trained using a batch size of 16 on a workstation equipped with an NVIDIA RTX 4070 Ti Super GPU. And optimized for a total of 500k training steps using the progressive three-stage training strategy described in Section \ref{training strategy}.

To evaluate deployment on resource-constrained hardware, the model is exported using ExecuTorch, and streaming inference is performed on a Raspberry Pi 3 B+. For all subsequent evaluations, the streaming model is configured with an input frame size of 3 and a decoder cache size of 5. As illustrated by the hyperparameter sweeps in Figure \ref{fig:hyperparameter_sweep}, audio quality improves with expanded context window with some diminishing returns. PESQ scores plateau as the input frame size increases beyond 2, whereas the computational cost (RTF) continues to scale. This specific configuration of 3 and 5 was selected because it provides an optimal balance, ensuring strict real-time execution constraints are met without sacrificing waveform reconstruction fidelity. Both speech reconstruction quality and real-time inference performance are measured under these conditions to assess the practicality of the proposed codec for edge-device voice communication.

\subsection{Baseline Comparison}
\vspace{-0.05in}
To properly contextualize the performance of the proposed codec, it must be evaluated through the lens of architectural efficiency. Current state-of-the-art neural audio codecs are designed with substantial parameter counts to maximize raw audio fidelity. To our knowledge, there are currently no established codecs in the literature engineered to operate at the ultra-low parameter footprint targeted by our work.

Table \ref{tab:baseline_comparison} presents an offline (non-streaming) comparison against these established models: Encodec~\cite{encodec}, DAC~\cite{dac}, and HILCodec~\cite{hilcodec}. Rather than strictly competing on raw upper-bound objective scores, the primary contribution of our model lies in bridging the gap between severe hardware constraints and high-fidelity audio, delivering highly competitive perceptual quality and intelligibility at a fraction of the computational and memory cost.

\begin{table}
\sisetup{
    group-separator = {,},
    group-minimum-digits = 4
}
\centering
\scriptsize
\caption{Baseline comparison of offline (non-streaming) reconstruction.}
\begin{tabular}{lcccc}
\hline
\textbf{Model}&
\textbf{Model Size} &
\textbf{Sample Rate} &
\textbf{PESQ $\uparrow$} &
\textbf{STOI $\uparrow$} \\
\hline
\textbf{\papername (ours)} & \num{773824} & 16khz & 1.3197 & 0.8105\\
Encodec& \num{14851810} & 24khz & 1.3209 & 0.8136\\
DAC & \num{74175906} & 16khz & 3.7507 & 0.9732\\
HILCodec & \num{9577019} & 24khz & 2.3953 & 0.9355\\
\hline
\end{tabular}
\label{tab:baseline_comparison}
%\vspace{-0.1in}
\end{table}

\subsection{Real Time Performance}
\begin{figure}[hbt]
    \centering
    \includegraphics[width=\linewidth]{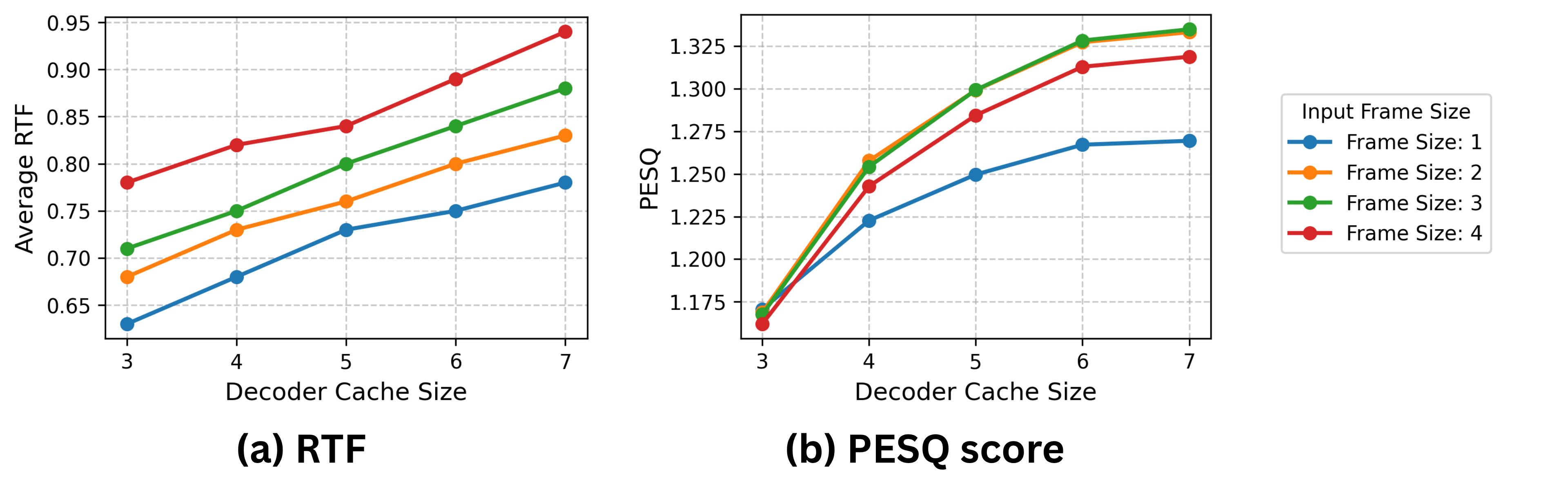}
    \vspace{-0.27in}
    \caption{Impact of streaming hyperparameters on inference efficiency and reconstruction fidelity. (a) Average Real-Time Factor (RTF) and (b) PESQ scores across various decoder cache and input frame sizes.}
    \label{fig:hyperparameter_sweep}
    \vspace{-0.1in}
\end{figure}

\vspace{-0.05in}
We evaluated the codec's real-time viability on a Raspberry Pi 3 Model B+ with model exported using PyTorch ExecuTorch~\cite{nachin2026executorchunifiedpytorch}. Efficiency is measured via the Real-Time Factor (RTF), where $\text{RTF} < 1.0$ is required for lag-free streaming. Figure \ref{fig:hyperparameter_sweep} maps the RTF landscape across varying input cache size and recon cache size configurations. While expanding these temporal buffers to capture wider context for the CNN layers increases the computational load, the optimized model maintains $\text{RTF} < 1.0$ across all  configurations. This allows to scale cache sizes flexibly without violating real-time edge constraints.

To demonstrate RFSQ's latency advantages, we profiled a standard RVQ baseline of our model (input frame size 3, decoder cache size 5). The RVQ variant yielded a prohibitive RTF of 3.09, violating the 20 ms frame budget. Specifically, the RVQ bottleneck alone consumed 48.50 ms, accounting for 78.7\% of the total 61.66 ms inference time. By eliminating the memory-bound codebook lookups that cause this bottleneck, RFSQ drastically reduces per-frame inference time and ensures real-time execution on resource-constrained microcontrollers.

\subsection{Ablation Study}
\vspace{-0.05in}
We evaluate perceptual quality and intelligibility using Perceptual Evaluation of Speech Quality (PESQ)~\cite{pesq}, and Speech Transmission Index for Intelligibility (STOI)~\cite{stoi}.

To further assess the preservation of high-level speech content, we compute the Semantic Difference (Semantic Diff), defined as one minus the cosine similarity between HuBERT~\cite{hsu2021hubertselfsupervisedspeechrepresentation} embeddings extracted from the reconstructed and reference audio:
\begin{equation}
    D_{\text{sem}} = 1 - \frac{\mathbf{h}_{\text{ref}} \cdot \mathbf{h}_{\text{recon}}}{\|\mathbf{h}_{\text{ref}}\|_2 \|\mathbf{h}_{\text{recon}}\|_2},
\end{equation}

where $\mathbf{h}_{\text{ref}}$ and $\mathbf{h}_{\text{recon}}$ denote the HuBERT embeddings of the reference and reconstructed audio signals, respectively. Unlike waveform-based metrics, this embedding-space measure captures semantic consistency and is less sensitive to minor signal-level variations. A lower Semantic Diff indicates that the reconstructed speech retains semantic information more faithfully.

Table \ref{tab:arch_comparison} presents the ablation study of key architectural and training components. The proposed streaming configuration achieves performance highly comparable to the offline upper bound, with only a marginal reduction in PESQ (1.3197 to 1.2994) and STOI (0.8105 to 0.8013), successfully validating the causal encoder and decoder adaptations.

The pseudo-lookahead strategy is critical for waveform fidelity; removing it degrades PESQ to 1.2600, confirming its necessity in mitigating iSTFT boundary artifacts. Similarly, ablating the MFCC loss causes the most severe performance drop (PESQ 1.1516) and more than doubles Semantic Difference (0.0699), indicating a substantial loss of speaker characteristics and structural speech features at extreme low bitrates.

Progressive training stabilizes the transition to discrete latents, improving overall waveform reconstruction. While the model trained without it achieves a nominally lower Semantic Difference (0.0317 vs. 0.0325), this marginal embedding-space improvement trades off with noticeable waveform-level degradation (PESQ drops to 1.2328), making progressive training a necessary compromise for signal fidelity.

Finally, replacing RFSQ with standard FSQ collapses reconstruction quality (PESQ 1.1266, STOI 0.7010), demonstrating that scale-invariant residual quantization is required to utilize the bit budget effectively. While the RVQ baseline yields the highest absolute scores, its memory-bound lookups violate real-time latency constraints (Section 4.3). RFSQ therefore provides the most practical compromise, delivering high intelligibility without exceeding edge computational budgets.

\begin{table}
\centering
\scriptsize
\caption{Encoder and deecoder architecture comparison (Checkpoint + Stream Variant Paired).}
\begin{tabular}{lcccc}
\hline
\textbf{Model} &
\textbf{PESQ $\uparrow$} &
\textbf{STOI $\uparrow$} &
\textbf{Semantic Diff. $\downarrow$} & \\
\hline
Offline (Non-streaming) & 1.3197 & 0.8105 & 0.0319 \\
\textbf{\papername (ours)} & \textbf{1.2994} & \textbf{0.8013} & 0.0325\\
\hline
w/o Pseudo-lookahead & 1.2600 & 0.7786 & 0.0329\\
w/o MFCC Loss & 1.1516 & 0.7328 & 0.0699\\
w/o Progressive Training & 1.2328 & 0.7888 & \textbf{0.0317}\\
RFSQ $\rightarrow$ FSQ & 1.1266 & 0.7010 & 0.0798\\
RFSQ $\rightarrow$ FSQ (w/o MFCC) & 1.1398 & 0.7133 & 0.0803\\
\hline
RFSQ $\rightarrow$ RVQ (Non-real-time) & 1.4214 & 0.8418 & 0.0238\\
\hline
\end{tabular}
\label{tab:arch_comparison}
%\vspace{-0.1in}
\end{table}
% \section{Discussion and Limitations}

% \subsubsection{Reconstruction Quality}
% To measure the reconstruction quality, we use Total Harmonic Distortion (THD). In the future, we plan to add support for more metrics such as MUSHRA, ViSQOL, STOI, MCD, and UTMOS.

% \section{Discussion}

% \subsection{Trade-offs Between Model Size and Quality}

% \subsection{Streaming Latency and Context}

% \subsection{Limitations}

% \subsection{Future Work}

\section{Conclusion}
We presented \papername, a lightweight streaming neural audio codec that achieves intelligible speech reconstruction at 2.3 kbps for resource-constrained edge devices. By pairing an RFSQ bottleneck with a pseudo-lookahead decoding framework, the system eliminates compute-intensive codebook searches and enables fully streamable iSTFT synthesis.
However, this pseudo-lookahead strategy inherently introduces a one-frame (20 ms) algorithmic delay into the communication pipeline. Furthermore, while the approach minimizes overlap-add artifacts, it does not completely mitigate boundary distortions.
Future development will focus on expanding the RFSQ bottleneck to support a variable bitrate (VBR) architecture that can dynamically adapt to fluctuating network throughput.

% ---------- References ----------
\clearpage
\bibliographystyle{IEEEtran}
\bibliography{reference}
\balance
\end{document}